\documentclass[12pt]{article}

\usepackage[margin=2.5cm]{geometry}

\usepackage{amsmath,amsthm,amsfonts,amscd,amssymb,bbm,mathrsfs,enumerate,url,bm}
\numberwithin{equation}{section}
\usepackage{authblk}

\usepackage{graphicx,tikz}
\usepackage[pdfborder={0 0 0}]{hyperref}
\usetikzlibrary{matrix,arrows}
\usetikzlibrary{decorations.pathreplacing}
\usetikzlibrary{decorations.pathmorphing}
\usetikzlibrary{patterns,fadings}
\allowdisplaybreaks

\def\semicolon{;}
\def\applytolist#1{
    \expandafter\def\csname multi#1\endcsname##1{
        \def\multiack{##1}\ifx\multiack\semicolon
            \def\next{\relax}
        \else
            \csname #1\endcsname{##1}
            \def\next{\csname multi#1\endcsname}
        \fi
        \next}
    \csname multi#1\endcsname}

\def\calc#1{\expandafter\def\csname c#1\endcsname{{\mathcal #1}}}
\applytolist{calc}QWERTYUIOPLKJHGFDSAZXCVBNM;
\def\bbc#1{\expandafter\def\csname bb#1\endcsname{{\mathbb #1}}}
\applytolist{bbc}QWERTYUIOPLKJHGFDSAZXCVBNM;
\def\bfc#1{\expandafter\def\csname bf#1\endcsname{{\mathbf #1}}}
\applytolist{bfc}QWERTYUIOPLKJHGFDSAZXCVBNM;
\def\sfc#1{\expandafter\def\csname s#1\endcsname{{\sf #1}}}
\applytolist{sfc}QWERTYUIOPLKJHGFDSAZXCVBNM;
\def\rfc#1{\expandafter\def\csname r#1\endcsname{{\mathrm #1}}}
\applytolist{rfc}QWERTYUIOPLKJHGFDSAZXCVBNM;
\def\scfc#1{\expandafter\def\csname sc#1\endcsname{{\mathscr #1}}}
\applytolist{scfc}QWERTYUIOPLKJHGFDSAZXCVBNM;

\usepackage{xspace}

\newcommand{\eps}{\varepsilon}

\def\bb1{\mathbbm{1}}

\def\<{\langle}
\def\>{\rangle}

\theoremstyle{remark}

\title{Some exact Green function solutions for\\[3pt]
non-linear classical field theories}

\author[1]{Marco Frasca\thanks{{\tt marcofrasca@mclink.it}}}

\affil[1]{Independent Researcher, 00176 Rome, Italy}

\author[2]{Stefan Groote\thanks{{\tt stefan.groote@ut.ee}}}
\affil[2]{F\"u\"usika Instituut, Tartu Ulikool,
  W.~Ostwaldi 1, EE-50411 Tartu, Estonia}

\date{ \ }

\begin{document}

\maketitle

\begin{abstract}
We consider some non-linear non-homogeneous partial differential equations
(PDEs) and derive their exact Green function solution as a functional Taylor
expansion in powers of the source. The kind of PDEs we consider are
dispersive ones where the exact solution of the corresponding homogeneous
equations can have some known shape. The technique has a formal similarity
with the Dyson--Schwinger set of equations to solve quantum field theories.
However, there are no physical constraints. Indeed, we show that a complete
coincidence with the statistical field model of a quartic scalar theory can be
achieved in the Gaussian expansion of the cumulants of the partition function.
\end{abstract}

\section{Introduction}

The availability of exact solutions to field theories, both classical and
quantum, is important to a general understanding of these theories. Such
solutions are very rare and in most of the cases are for unphysical situations.
The harnessing of the functional renormalization group can help to mitigate
such a limitation significantly~\cite{Polchinski:1983gv,Wetterich:1992yh,%
Wipf:2021mns}. Still, explicit solutions are not provided in this way.

Recently, an exact solution to some quantum field theory was
proposed~\cite{Frasca:2015wva,Frasca:2015yva}. This means that all the
correlation functions of the theory could in principle be evaluated exactly.
Besides this, explicit solutions were provided for the one-point (1P) and
two-point (2P) correlation functions by using the properties of the Jacobian
elliptic functions~\cite{Bateman:1953gv}. The appearance of such
transcendental functions arises by the very nature of the partial differential
equations (PDEs) that characterize their equations of motion, mirrored into
their quantum counterpart. These kind of exact solutions are obtained assuming
a specific property of the higher-order correlation functions that are taken
to be zero when evaluated to identical arguments. This yields all of them fully
given by the 1P- and 2P-correlation functions, typically in a Gaussian type.

The idea can be traced back to recently proposed exact solutions obtained for
nonlinear PDEs~\cite{Frasca:2009bc}. We show that this set of solutions is
very well mapped to the solutions of the corresponding classical counterparts
in a functional Taylor expansion in the source. In Ref.~\cite{Frasca:2013tma}
we have shown that this series can correspond to a strong coupling expansion in
the inverse of the coupling.

The aim of this paper is to give a sound mathematical understanding of such
solutions providing relevant new approaches to re-derive the main results. We
also yield a complete solution to a quantum SU(2) Yang--Mills theory providing
explicitly a proof of the Gaussian nature of the kind of solution we obtained.
This means that in principle we are able to evaluate all the averages in
closed form, even though there could be no need to go to very high orders. For
instance, in our case we have a Gaussian distribution and we can limit our
analysis to the mean and the covariance even if all higher order cumulants
could in principle be evaluated. 

The main reason to get a closed-form solution arises from studies on the
lattice of the gluon and ghost propagators for the Yang--Mills theory, mostly
in the Landau gauge~\cite{Bogolubsky:2007ud,Cucchieri:2007md,Oliveira:2007px},
and the corresponding spectrum~\cite{Lucini:2004my,Chen:2005mg} that shows
unequivocally that a mass gap appears also when the interaction with fermions
is neglected. Earlier theoretical analysis supported these
results~\cite{Cornwall:1981zr,Cornwall:2010bk,Dudal:2008sp,Frasca:2007uz,%
Frasca:2009yp,Frasca:2015yva} by providing closed-form formulas for the gluon
propagator. Quite recently, the set of Dyson--Schwinger equations for this
case was solved, for the 1P- and 2P-correlation functions, and the spectrum
was computed very accurately both in three and four dimensions
\cite{Frasca:2016sky,Frasca:2015yva,Frasca:2017slg}, proving also confinement
as a property of the theory~\cite{Frasca:2016sky,Chaichian:2018cyv}. These
latter results are strongly linked to the solution of the quartic scalar field
theory mapped onto the Yang--Mills theory, as shown in
Refs.~\cite{Frasca:2009yp,Frasca:2015yva}.

Indeed, our studies rely on Dyson--Schwinger equations in a PDE shape that
appears to be the most sensible approach to treat a non-perturbative quantum
field theory~\cite{Baker:1976vz,Eichten:1974et,Roberts:1994dr}. Indeed,
Bender, Milton and Savage~\cite{Bender:1999ek} proposed to derive the
Dyson--Schwinger equations and treat them in differential form. This way to
manage these equations was the one used to find the exact
solution~\cite{Frasca:2015yva}. The approach turns out to be quite useful when
a non-trivial solution of the equation for the 1P-correlation function, or the
equation of motion, is known. Therefore, a complete solution to theories that
normally are considered treatable only through perturbation methods can become
available.

The paper is structured as follows. In Sec.~\ref{prep} we introduce the
theories we aim to analyze. In Sec.~\ref{scalar_field} we solve the classical
theory of a massless quartic scalar field. In Sec.~\ref{Y-M_field} we solve the
classical equations for the SU(2) Yang--Mills theory. In Sec.~\ref{part_func}
we show how such classical solutions map onto the corresponding statistical
field theory of the scalar field. In Sec.~\ref{conc} we present our
conclusions.

\section{Preliminaries}\label{prep}

We consider a scalar field $\phi$ with a single component and $Z_2$ symmetry
with the action
\begin{equation}
S_\phi=\int d^4x\left[\frac12(\partial\phi)^2-\frac\lambda4\phi^4+j\phi\right],
\end{equation}
where $j$ is an arbitrary source. Our aim is to solve the equation of motion
\begin{equation}\label{eq:phi}
\partial^2\phi+\lambda\phi^3=j,
\end{equation}
where the constant $\lambda$ is kept to obtain an understanding of the
solutions on the strength of the given coupling. Working in $d=4$ dimensions
has the advantage to keep such a constant dimensionless. We will show that
this model can be completely solved using elliptic functions. Similarly, we
evaluate the corresponding partition function
\begin{equation}
    Z[j]=\int[d\phi]e^{-S_\phi},
\end{equation}
showing how the classical solution extends to the quantum field theory through
the evaluation of the correlation functions. The structure is of Gaussian type.

For the Yang--Mills theory, we limit our analysis to the SU(2) group. We will
have the action
\begin{equation}
S_A=\int d^4x\left(-\frac14F\cdot F+j\cdot A\right)
\end{equation}
where $j_\mu$ is a generic source and an element of the su(2) algebra. The
equation of motion is given by (Latin letters $a,b,c,\ldots$ represent group
indices taking the values $1,2,3$)
\begin{equation}
F_{\mu\nu}^a=\partial_\mu A_\nu^a-\partial_\nu A_\mu^a
  +g\epsilon^{abc}A_\mu^b A_\nu^c
\end{equation}
where $g$ plays the same role as $\lambda$ before, leading to the equations of
motion
\begin{equation}\label{eq:YM}
\partial^\mu F_{\mu\nu}^a+g\epsilon^{abc}A^{b\mu}F^c_{\mu\nu}=j_\nu^a,
\end{equation}
where $\epsilon^{abc}=1$ for even permutations of the indexes $abc$,
$\epsilon^{abc}=-1$ for odd permutations, and $\epsilon^{abc}=0$ if two or
more indexes are equal. In the following we will provide a solution to
Eq.~(\ref{eq:YM}). We will show how the solution can be obtained through
the equations of the scalar field given a mapping theorem between these two
theories~\cite{Frasca:2007uz,Frasca:2009yp}.

\section{Solution for the scalar field}\label{scalar_field}

\subsection{Homogeneous equation}

Our ability to solve the full problem relies heavily on the solution of the
homogeneous equation
\begin{equation}
\partial^2\phi_0+\lambda\phi_0^3=0.
\end{equation}
To solve this equation, we introduce the wave-like coordinate $\xi=p\cdot x$
where $p$ is the momentum four-vector. In terms of this coordinate we have
\begin{equation}
p^2\frac{d^2\phi_0(\xi)}{d\xi^2}+\lambda\phi_0^3(\xi)=0.
\end{equation}
The solution of this equation can be easily written down as
\begin{equation}
\phi_0(\xi)=a\operatorname{sn}(\xi+\theta,i)
\end{equation}
with $a^2\lambda/p^2=2$, where $a$ and $\theta$ are integration constants and
$\operatorname{sn}(z,i)$ is Jacobi's elliptic sine of elliptic modulus $i$.
Due to the extensive use in literature, we fix $a^2=\sqrt{2/\lambda}\mu^2$ and
are left with the dispersion relation
\begin{equation}\label{eq:disp}
p^2=\mu^2\sqrt{\frac{\lambda}{2}}.
\end{equation}
Therefore, our solution of the homogeneous equation takes the form
\begin{equation}
\phi_0(x)=\mu\left(\frac{2}{\lambda}\right)^{1/4}
  \operatorname{sn}(p\cdot x+\theta,i).
\end{equation}
From a physical point of view, the solution represents a non-linear wave
moving at a speed $v=|{\bm p}|/p_0<1$ with dispersion relation provided by
Eq.~(\ref{eq:disp}).

We note that the technique we used to get our solution is similar to that
applied to obtain the Volkov solution of the Dirac
equation~\cite{Wolkow:1935zz}. The kind of solution we have obtained is indeed
a Fubini--Lipatov instanton~\cite{Fubini:1976jm,Lipatov:1976ny}, as pointed
out in our preceding works (see Ref.~\cite{Frasca:2023qii} and references
therein) and firstly noted in Ref.~\cite{Frasca:2011bd} for the ground state
of the Yang--Mills theory, quite different in its physical interpretation to
the Volkov solution. Anyway, the choice of this solution is arbitrary. In our
case, it grants an exact analytical solution for the given equation and for
the Green function which is all we need for our aims.

\subsection{Functional Taylor expansion}

In order to find a general solution, we assume that the solution is a
functional of the source $j$, $\phi=\phi[j]$. This functional is expanded in a
functional Taylor series
\begin{equation}
\phi[j]=\phi_0(x)+\sum_{k=1}^\infty\int
  C_k(x,x_1,\ldots,x_k)\prod_{l=1}^kj(x_l)d^4x_l.
\end{equation}
Our aim will be to obtain the Green functions $C_k(x_1,\ldots,x_k)$, the
analogous of the correlation functions in statistical field theory, in closed
form. It is not difficult to see that
\begin{equation}
C_k(x,x_1,\dots,x_k)=\frac{\delta^k\phi[j]}{\delta j(x_1)\ldots\delta j(x_k)}.
\end{equation}
The general solution $\phi[j]$ satisfies the equation of motion~(\ref{eq:phi}).
Accordingly, for the Green functions $C_k$ at vanishing source $j=0$ we obtain
the set of equations
\begin{eqnarray}
\lefteqn{\partial^2\phi_0(x)+\lambda\phi_0^3(x)=0,}\\[12pt]
\lefteqn{\partial^2C_1(x,x_1)+3\lambda\phi_0^2(x)C_1(x,x_1)
  =\delta^4(x-x_1),}\\[12pt]
\lefteqn{\partial^2C_2(x,x_1,x_2)+3\lambda\phi_0^2(x)C_2(x,x_1,x_2)
  +6\lambda\phi_0(x)C_1(x,x_1)C_1(x,x_2)=0,}\\[12pt]
\lefteqn{\partial^2C_3(x,x_1,x_2,x_3)+3\lambda\phi_0^2(x)C_3(x,x_1,x_2,x_3)}
  \nonumber\\&&\strut
  +6\lambda\phi_0(x)C_1(x,x_3)C_2(x,x_1,x_2)
  +6\lambda C_1(x,x_1)C_1(x,x_2)C_1(x,x_3)\nonumber\\&&\strut
  +6\lambda\phi_0(x)C_2(x,x_1,x_3)C_1(x,x_2)
  +6\lambda\phi_0(x)C_1(x,x_1)C_2(x,x_2,x_3) = 0,\quad\ldots\qquad
\end{eqnarray}
From this set of equations we easily recognize that $C_1(x,x_1)$ is the Green
function of this non-linear problem. The corresponding PDE is linear and,
therefore, in principle amenable to an analytic solution. It is also clear
that the functions $\{C_k|k>1\}$ are all given by combinations of $\phi_0$ and
$C_1$. Therefore, the problem is completely solved if we are able to get a
closed-form solution for these two functions. We already know $\phi_0$. Thus,
our aim is to determine $C_1$.

It is interesting to note the relevant difference to the linear case. In such
a case, all the functions $C_n$ with $n>1$ are 0. Thus, it is easy to conclude
that nonlinear wave propagation puts in existence all these higher order Green
functions. This is very similar to the application of the Fourier series to a
nonlinear wave equation that gives rise to higher order harmonics from a
single mode plane wave. Higher order correlation functions acquire a
well-defined meaning in a statistical sense as we will see below.

\subsection{Green function}

We have to solve the equation
\begin{equation}
    \partial^2C_1(x,x_1)+3\lambda\phi_0^2(x)C_1(x,x_1)=\delta^4(x-x_1),
\end{equation}
where
\begin{equation}
\phi_0(x)=\mu\left(\frac{2}{\lambda}\right)^{1/4}
  \operatorname{sn}(p\cdot x+\theta,i).
\end{equation}
It is easy to write down a solution for the associate homogeneous equation
given by
\begin{equation}
\phi_h(x)=a\operatorname{cn}(p\cdot x+\theta,i)
  \operatorname{dn}(p\cdot x+\theta,i),
\end{equation}
where $\operatorname{cn}$ and $\operatorname{dn}$ are Jacobian elliptic
functions of elliptic modulus $i$. Observing that
\begin{equation}
\frac{d}{dz}\operatorname{sn}(z,i)=\operatorname{cn}(z,i)\operatorname{dn}(z,i)
\end{equation}
and
\begin{equation}
\operatorname{sn}(z,i)=\frac{2\pi}{K(i)}\sum_{n=0}^\infty(-1)^n
  \frac{e^{-\left(n+\frac12\right)\pi}}{1+e^{-(2n+1)\pi}}
  \sin\left((2n+1)\frac{\pi z}{2K(i)}\right),
\end{equation}
the solution of the homogeneous solution can be written as a Fourier series,
\begin{equation}
\phi_h(x)=\frac{\pi^2}{K^2(i)}\sum_{n=0}^\infty(-1)^n(2n+1)
  \frac{e^{-\left(n+\frac12\right)\pi}}{1+e^{-(2n+1)\pi}}
  \cos\left((2n+1)\frac{\pi}{2K(i)}(p\cdot x+\theta)\right).
\end{equation}
The choice $\theta=(4m+1)K(i)$ with $m\in\mathbb{Z}$ for the phase grants that
the function $\phi_0(x)$ is zero on the light cone. In quantum field theory
this solution would immediately give the Feynman propagator. In classical
field theory, we have to use the Laplace--Fourier transform to get the initial
conditions right. We want to preserve the causal behaviour of the Feynman
propagator in quantum field theory by fixing the boundary conditions in such a
way that positive energy solutions propagate forward in time and negative
energy solutions propagate backward in time, as our aim is to put at work this
Green function in a fully quantum formulation of the theory.
In order to work out this aim, we calculate
\begin{equation}
C_1(\omega,{\bm x})=\int_0^\infty dt e^{-i(\omega+i\eps)t}\phi_h(x),
\end{equation}
where $\eps>0$ is a small quantity which is sent to zero in the end in order
to obtain the correct (Feynman) integration path. The integration gives
\begin{equation}
C_1(\omega,{\bm x})=\frac12\sum_{n=0}^\infty A_n
  \left[\frac{e^{i(2n+1)\frac{\pi}{2K(i)}{\bm p}\cdot{\bm x}}}{(\omega
  -(2n+1)\frac{\pi}{2K(i)}p_0)+i\eps}-\frac{e^{-i(2n+1)\frac{\pi}{2K(i)}{\bm p}
  \cdot{\bm x}}}{(\omega+(2n+1)\frac{\pi}{2K(i)}p_0)+i\eps}\right],
\end{equation}
where
\begin{equation}
A_n=\frac{\pi^2}{K^2(i)}(2n+1)
  \frac{e^{-\left(n+\frac{1}{2}\right)\pi}}{1+e^{-(2n+1)\pi}}.
\end{equation}
We can perform a Fourier transform in space and use the condition $p^2=m^2$ to
obtain
\begin{eqnarray}
C_1(k)&=&\frac12\sum_{n=0}^\infty A_n
  \Bigg[\frac{\delta^3({\bm k}+(2n+1)\frac{\pi}{2K(i)}{\bm p})}{(\omega
  -(2n+1)\frac{\pi}{2K(i)}\sqrt{{\bm p}^2+m^2})+i\eps}\nonumber\\&&\strut
  -\frac{\delta^3({\bm k}-(2n+1)\frac{\pi}{2K(i)}{\bm p})}{(\omega
  +(2n+1)\frac{\pi}{2K(i)}\sqrt{{\bm p}^2+m^2}))+i\eps}\Bigg].
\end{eqnarray}
Finally, we integrate out the arbitrary three-momentum $p$ in order to get to
the Feynman propagator (which is explicitly given only in full momentum
space), using the covariant integration
$\int d^3p/2E_p$ with $E_p=\sqrt{{\bm p}^2+m^2}$, to obtain
\begin{eqnarray}
C_1(k)&=&\frac{1}{4\sqrt{{\bm k}^2/((2n+1)\frac{\pi}{2K(i)})^2+m^2}}
  \sum_{n=0}^\infty A_n\times\strut\nonumber\\&&\strut
  \Bigg[\frac{1}{(\omega-(2n+1)\frac{\pi}{2K(i)}\sqrt{{\bm k}^2/((2n+1)
    \frac{\pi}{2K(i)})^2+m^2})+i\eps}\nonumber\\&&\strut
  -\frac{1}{(\omega+(2n+1)\frac{\pi}{2K(i)}\sqrt{{\bm k}^2/((2n+1)
    \frac{\pi}{2K(i)})^2+m^2})+i\eps}\Bigg].
\end{eqnarray}
For the Feynman propagator we have
\begin{equation}
G(x-y)=\frac{1}{4\pi}\delta(\tau_{xy}^2)+\phi_h(x-y),
\end{equation}
where $\tau_{xy}^2=(x_0-y_0)^2-(x_1-y_1)^2-(x_2-y_2)^2-(x_3-y_3)^2$.
Using\footnote{A simple verification of this result, normally not reported in
textbooks, can be found at e.g.\
\url{https://physics.stackexchange.com/q/615102}.}
$\Box\delta(\tau_{xy}^2)=4\pi\delta^4(x-y)$ and our choice of $\theta$ such
that $\operatorname{sn}^2(p\cdot(x-y)+\theta,i)\delta(\tau_{xy}^2)
=\delta(\tau_{xy}^2)$, that is zero outside of the light cone, we can indeed
show that this propagator solves the PDE for the Green function. As both
$G(x-y)$ and $C_1(x,y)$ are solving the same PDE, we can conclude that
$C_1(x,y)=G(x-y)$.

\subsection{Higher order Green functions}

Having calculated the Green function $C_1$ makes it easy to get all the higher
order Green functions. For completeness we can write a couple of them as
\begin{eqnarray}
\lefteqn{C_2(x,y,z)=-6\lambda\int d^4wC_1(x,w)\phi_0(w)C_1(w,y)C_1(w,z),}
  \\[12pt]
\lefteqn{C_3(x,y,z,w)=-6\lambda\int d^4vC_1(x,v)\phi_0(v)C_1(v,y)C_2(v,z,w)}
  \nonumber\\&&\strut
  -6\lambda\int d^4v C_1(x,v)C_1(v,y)C_1(v,z)C_1(v,w)
  -6\lambda\int d^4vC_1(x,v)\phi_0(v)C_2(v,y,z)C_1(v,w)\nonumber\\&&\strut
  -6\lambda\int d^4vC_1(x,v)\phi_0(v)C_1(v,y)C_2(v,z,w),\quad\ldots\qquad
\end{eqnarray}
As previously stated, these higher order Green functions are obtained by
combination of $\phi_0$ and $C_1$ and nothing else. Therefore, we conclude
that the problem is completely solved by knowing these latter functions.

\section{Solution for the Yang-Mills field}\label{Y-M_field}

If we consider the current expansion
\begin{equation}
A_\mu^{a}[j]=A_\mu^{a(0)}(x)+\sum_{k=1}^\infty\int
  C_{\mu_1\ldots \mu_k}^{a_1\ldots a_k(k)}(x,x_1,\ldots,x_k)
  \prod_{l=1}^kj^{a_l\mu_l}(x_l)d^4x_l,
\end{equation}
the zeroth order term solves exactly the homogeneous equation
\begin{eqnarray}
\lefteqn{\partial^\mu(\partial_\mu A_\nu^{a(0)}-\partial_\nu A_\mu^{a(0)}
  +g\epsilon^{abc}A_\mu^{b(0)}A_\nu^{c(0)})}\nonumber\\&&\strut
  +g\epsilon^{abc}A^{b(0)\mu}(\partial_\mu A_\nu^{c(0)}
  -\partial_\nu A_\mu^{c(0)}+g\epsilon^{cde}A_\mu^{d(0)}A_\nu^{e(0)})=0,\qquad
\end{eqnarray}
once we select the Lorenz gauge \cite{Frasca:2015yva}. The connection to the
specific set of solutions can be accomplished by introducing the mixed symbols
for SU(2) ($a=1,2,3$ is the group index)
\begin{equation}\label{eq:msym}
\eta_\mu^1=(0,1,0,0), \quad \eta_\mu^2=(0,0,1,0), \quad \eta_\mu^3=(0,0,0,1),
\end{equation}
and using a mapping between a scalar field and the Yang--Mills
theory~\cite{Frasca:2009yp}
\begin{equation}
A_\mu^{a(0)}(x)=\eta_\mu^a\phi_0(x),
\end{equation}
where $\phi_0(x)$ is the solution provided for the scalar field. By
calculating the variation with respect to $j_\rho^f$, we derive the equations
of motion~(\ref{eq:YM}), in Lorenz gauge yielding
\begin{eqnarray}
\lefteqn{\partial^2C_{\nu\rho}^{af(1)}(x,y)
  +g\epsilon^{abc}C_{\mu\rho}^{bf(1)}(x,y)\partial^\mu A_\nu^{c(0)}(x)
    +g\epsilon^{abc}A_\mu^{b(0)}(x)\partial^\mu C_{\nu\rho}^{cf(1)}(x,y)}
    \nonumber\\&&\strut
    +g\epsilon^{abc}C^{bf(1)\mu}_\rho(x,y)\partial_\mu A_\nu^{c(0)}(x)
    -g\epsilon^{abc}A^{b(0)\mu}(x)\partial_\nu C_{\mu\rho}^{cf(1)}(x,y)
    \nonumber\\&&\strut
    +g^2\epsilon^{abc}\epsilon^{cde}C_\rho^{bf(1)\mu}(x,y)A_\mu^{d(0)}(x)
    A_\nu^{e(0)}(x)+g^2\epsilon^{abc}\epsilon^{cde}A^{b(0)\mu}(x)
    C_{\mu\rho}^{df(1)}(x,y)A_\nu^{e(0)}(x)\nonumber\\&&\strut
    +g^2\epsilon^{abc}\epsilon^{cde}A^{b(0)\mu}(x)A_\mu^{d(0)}(x)
    C_{\nu\rho}^{ef(1)}(x,y)=\eta_{\nu\rho}\delta_{af}\delta^4(x-y).\qquad
\end{eqnarray}
Using the mapping theorem for SU(2), this equation collapses to
\begin{equation}
\partial^2C_1(x,y)+6g^2\phi_0^2(x)C_1(x,y)=\delta^4(x-y),
\end{equation}
and we are back to the scalar field case provided we choose $\lambda=2g^2$,
the factor $2$ being a Casimir operator of SU(2). Therefore, from this point
on everything is obtained in an identical way. Just in order to complete our
derivation, we write down the final formula for the classical SU(2)
Yang--Mills theory in Lorenz gauge,
\begin{equation}
C_{\mu\nu}^{ab}(k)=\delta_{ab}\left(\eta_{\mu\nu}
  -\frac{k_\mu k_\nu}{k^2}\right)C_1(k),
\end{equation}
where $C_1(k)$ is the Green function we computed for the scalar field theory.
We emphasize that sets of mixed symbols like those given in Eq.~(\ref{eq:msym})
can be provided in principle for any gauge group, making this kind of
classical solution easily generalised.

\section{Partition function of the scalar field}\label{part_func}

We have seen that we are able to solve exactly some non-linear classical
theory provided we are able to compute the solution of the homogeneous
equation and the Green function of the first order. A functional series
expansion gives the solution one is looking for at any desired order. This way
to find exact solutions is very similar to what is done in statistical or
quantum field theory. If the whole set of correlation functions is computable
in closed form at any order one could claim in principle to have solved a
given theory exactly. In most common applications, generally provided by
realistic models that are mostly given by non-linear PDEs, this is not
possible except for the free case. Therefore, one could ask how to extend the
above exact solution for the scalar field to statistical field theory. In
order to understand this, let us consider a generic scalar field theory
having a partition function
\begin{equation}
Z[j]=\int[d\phi]e^{-S[\phi]-\int j\phi d^4x}.
\end{equation}
We will have a set of Schwinger functions
\begin{equation}
S_n(x_1,\ldots,x_n)=\frac{1}{Z}\frac{\delta^nZ}{\delta j(x_1)\ldots
  \delta j(x_n)}\Big|_{j=0}=Z^{-1}\int[d\phi]\phi(x_1)\ldots\phi(x_n)
    e^{-S[\phi]-\int j\phi d^4x}\Big|_{j=0}.
\end{equation}
These functions yield the connected correlation functions
\begin{equation}
G_n(x_1,\ldots,x_n)=\frac{\delta^n\ln Z}{\delta j(x_1)\ldots\delta j(x_n)}
  \Big|_{j=0}.
\end{equation}
The relations
\begin{eqnarray}
S_1(x_1)&=&G_1(x_1),\nonumber\\
S_2(x_1,x_2)&=&G_2(x_1,x_2)+G_1(x_1)G_1(x_2),\nonumber\\
S_3(x_1,x_2,x_3)&=&G_3(x_1,x_2,x_3)+G_2(x_1,x_2)G_1(x_3)+G_2(x_1,x_3)G_1(x_2)
  \nonumber\\&&\strut
  +G_2(x_2,x_3)G_1(x_1)+G_1(x_1)G_1(x_2)G_1(x_3),\quad\ldots\qquad
\end{eqnarray}
hold between Schwinger and connected functions. We say that a theory will
admit a Gaussian solution if all the connected functions (cumulants) of order
higher than $2$ can be computed through the cumulants of order $1$ and $2$.
This is very well known from statistics. These Gaussian solutions completely
map onto those we have found for the classical theories. A trivial example of
a Gaussian solution is found in a free massive scalar theory. In our case, we
aim to provide some non-trivial examples.

If we do not care too much about translation invariance that might or might
not be broken by nature, a straightforward way to obtain such kind of Gaussian
solutions for a given statistical field theory is by solving the
Dyson--Schwinger set of equations, assuming that higher-order connected
correlation functions ($n>2$) are zero if two or more arguments coincide. In
such a case, we are in principle able to completely solve the set of
equations, as shown in Ref.~\cite{Frasca:2015yva}, because each equation
becomes independent from the higher order ones in the set. This class of
Gaussian solutions of the Dyson--Schwinger set of equations for a statistical
field theory could prove to be relevant in physics in case that agreement with
lattice and/or experimental data is achieved. In all other cases, it is just a
technique to get some closed-form analytical solutions, having significant
pedagogical meaning also for realistic theories where such solutions are
currently missing. The solutions of the Dyson--Schwinger equations completely
map the classical solutions. 

For the sake of completeness, we list here the first few Dyson--Schwinger
equations for a massless quartic scalar theory~\cite{Frasca:2015yva}, showing
how these equations match the classical equations we presented before.
\begin{eqnarray}
\lefteqn{\partial^2 G_1(x)+\lambda\Big[G_1^3(x)+3G_2(x,x)G_1(x)+G_3(x,x,x)
  \Big]=0,}\\[12pt]
\lefteqn{\partial^2G_2(x,y)+\lambda\Big[3(G_1^2(x)G_2(x,y)+3G_2(x,x)G_2(x,y)}
  \nonumber\\&&\strut
    +3G_3(x,x,y)G_1(x)+G_4(x,x,x,y)\Big]=\delta^4(x-y),\\[12pt]
\lefteqn{\partial^2G_3(x,y,z)+\lambda\Big[6G_1(x)G_2(x,y)G_2(x,z)
  +3G_1^2(x)G_3(x,y,z)}\nonumber\\&&\strut
  +3G_2(x,z)G_3(x,x,y)+3G_2(x,y)G_3(x,x,z)+3G_2(x,x)G_3(x,y,z)
  \nonumber\\&&\strut
  +3G_1(x)G_4(x,x,y,z)+G_5(x,x,x,y,z)\Big]=0,\\[12pt]
\lefteqn{\partial^2G_4(x,y,z,w)+\lambda\Big[6G_2(x,y)G_2(x,z)G_2(x,w)
  +6G_1(x)G_2(x,y)G_3(x,z,w)}\nonumber\\&&\strut
  +6G_1(x)G_2(x,z)G_3(x,y,w)+6G_1(x)G_2(x,w)G_3(x,y,z)+3G_1^2(x)G_4(x,y,z,w)
  \nonumber\\&&\strut
  +3G_2(x,y)G_4(x,x,z,w)+3G_2(x,z)G_4(x,x,y,w)+3G_2(x,w)G_4(x,x,y,z)
  \nonumber\\&&\strut
  +3G_2(x,x)G_4(x,y,z,w)+3G_1(x)G_5(x,x,y,z,w)+G_6(x,x,x,y,z,w)\Big]=0,
  \quad\ldots\qquad
\end{eqnarray}
It is easy to see that except for the quantum correction arising from
$G_2(x,x)$, the choice $\{G_k(x,x,\ldots)=0|k>2\}$ completely maps the above
set to the classical solution. In quantum field theory the quantum correction
entails some renormalization technique to be evaluated. The final result is
consistent and yields a Gaussian solution for the theory.

It should be pointed out that non-Gaussian solutions are also possible, even
if they are not exact like this one. This crucially depends on the choice of
the ground state solution of $G_1$, that is the choice of the solutions of the
PDE for the 1P-correlation function. A well-known example is the perturbative
solution that chooses $G_1$ being a constant. Our exact solution can also be
taken as an unperturbed starting solution to evaluate further non-Gaussian
corrections as for quantum chromodynamics.

\section{Conclusions}\label{conc}

We have provided exact Green function solutions to classical nonlinear field
theories and have shown that such solutions map very well onto the
corresponding analysis through Dyson--Schwinger equations in statistical field
theory. This special solutions represent non-trivial Gaussian solutions for
the latter. It is enough to know the first two correlation functions (the
exact solution of the homogeneous equation and the Green function for the
classical theory) to perform in principle whatever computation in the given
theory. This also provides a generic way to perform computations in classical
field theory in case that nonlinear PDEs are involved and some solutions of
the corresponding homogeneous equations are known.

An interesting field of applications for this technique would be black hole
physics~\cite{Gussmann:2016mkp,Guo:2020caw,Al-Badawi:2021wdm}. Indeed,
something in this direction has been done by one of us
(M.F.)~\cite{Frasca:2005fs,Frasca:2020jzp}. We will exploit this application
further in our future works.

\end{document}